\documentclass[aps,prl,twocolumn,floatfix,showpacs]{revtex4}
\usepackage{amsmath}
\usepackage{amssymb}
\usepackage[dvips]{graphicx}

\newcommand{\up}{\uparrow}
\newcommand{\down}{\downarrow}

\begin{document}

\title{Luttinger liquid of trimers in  Fermi gases with unequal masses}

\author{Giuliano Orso}
\author{Evgeni Burovski}
\author{Thierry Jolicoeur}

\affiliation{Laboratoire de Physique Th\'eorique et Mod\`eles statistiques,
Universit\'e Paris-Sud, 91405 Orsay, France}

%

\begin{abstract}
We investigate one dimensional attractive Fermi gases in spin-dependent optical lattices.
We show that three-body bound states - ``trimers'' - exist as soon as the two tunneling rates
are different. We calculate the binding energy and the effective mass of a single trimer.
We then show numerically that for finite and commensurate densities $n_\uparrow=n_\down/2$ an energy gap appears, 
implying that the gas is a one-component Luttinger liquid of trimers with suppressed superfluid correlations. 
The boundaries of this novel phase are given. We discuss experimental situations to test our predictions.


\end{abstract}


\maketitle

Recent advances with ultra-cold atoms are opening new prospects to address
fundamental theoretical issues in direct experiments \cite{PitaevskiiRMP2008}.
A long standing problem  is whether superconductivity can coexist with the
presence of an unequal number of up and down fermions. 
An intriguing possibility is the celebrated Fulde-Ferrell-Larkin-Ovchinnikov (FFLO) state \cite{FFLO}, where the superconducting order parameter becomes modulated in space. 
The experimental search for  polarized superfluids in atomic quantum
gases has so far been restricted to 3D configurations \cite{mit,rice}.
A new and promising direction is to confine atoms in higly elongated traps, where the 
FFLO state is known to be very robust \cite{yang, Orso2007,Drummond2007}, as  confirmed by detailed numerical simulations \cite{Feiguin}.

Another exciting topic that is currently being explored experimentally is the
pairing in Fermi gases with unequal masses, like mixtures of $^6$Li and $^{40}$K near a heteronuclear Feshbach resonance
\cite{Wille2008,dieckmann, unequal}. 
Alternatively, one can also trap a two-component Fermi gas in a spin dependent optical lattice so that the corresponding 
effective masses are different \cite{mandel}. 
%
Assuming that the transverse motion of atoms is frozen by a strong radial confinement, the system can then be described by the 
1D asymmetric Fermi-Hubbard~\cite{GiamarchiCazalillaHo2005,Batrouni2009,DasSarma2009} model~:
\begin{equation}
H_\mathrm{aH} = -\sum_{\langle ij \rangle\sigma} t_\sigma \left(
c^\dagger_{i,\sigma} c_{j,\sigma} +h.c. \right) +
U\sum_{i}\hat{n}_{i\uparrow}\hat{n}_{i\downarrow} \; ,
\label{aHubb}
\end{equation}
where $U<0$ is the on-site attraction and $t_\sigma$ are the spin-dependent
tunneling rates. Here $c_{i\sigma}$ annihilates a fermion with spin $\sigma$
at site $i$  and $\hat{n}_{i\sigma}$ is the local density.

For $t_\up=t_\down$ Eq.\ \eqref{aHubb} reduces to the Hubbard model.
For $t_\down=0$ Eq.\ \eqref{aHubb} is nothing but a spinless version of 
the Falicov-Kimball model \cite{FK1969}, originally devised for explaining
metal-insulator transition in mixed-valence materials via hybridization
of itinerant electrons ($\up$) and localized holes ($\down$). 
There is a devil's staircase structure in the ground state of the system \cite{Yeomans1997}.
For finite $t_\down$
Eq.\ \eqref{aHubb} is often referred to as an extended Falicov-Kimball model, 
which has been recently suggested to feature Bose-Einstein condensation
of d-f excitons with macroscopic polarization \cite{FK1996}
and a spontaneous ferroelectric state \cite{FK2002}.

For equal masses, $t_\down=t_\uparrow$, the exact (Bethe ansatz) solution  shows
that $n$-body bound states with $n>2$ are  generally forbidden \cite{takahashi}. 
When the tunneling rates are different, $t_\downarrow<t_\uparrow$, the above model is no longer integrable and many interesting questions arise: are three-body bound states (trimers) allowed ? What are their properties ? Can trimers open an energy gap like pairs do ? And if so, what are the differences between 
the two gapped phases ?
The purpose of this Letter is to provide an explicit answer to these relevant questions. 
We show that there is formation
of three-body bound states of two heavy ($\down$) fermions and one light ($\up$) fermion.
We first find the regime where these  trimers are stable as a function of mass asymmetry
and attraction strength by solving the three-body problem. We then use the DMRG method to show that
at low but finite density  there exists a novel phase with a nonzero energy gap which is a one-component 
Luttinger liquid of trimers, with \textsl{exponentially} suppressed  superconducting FFLO correlations. 
Finally, we calculate the boundaries of the trimer phase in the grand canonical phase diagram.
These results are in agreement with the generic bosonization
analysis of Ref.~\cite{bosonization} where the role of higher harmonics of the density operator was
elucidated. 
The trimers discussed here are a cold atom analog of the trions recently
observed in semiconductors~\cite{trion}. Three-body bound states (though of different origin)
have also been predicted to occur in three-component Fermi gases \cite{arazia}.
In the following we fix the energy scale in Eq.\ (\ref{aHubb}) by setting $t_\up$=1  and assume $t_\down \leqslant 1$ without loss
of generality.


\noindent\textit{Three-body problem.---} We start by calculating the binding
energy and the effective mass of the trimer. The Schr\"{o}dinger equation can
be conveniently rewritten in  integral form by using Green functions. 
For the three-body problem in a lattice one finds~\cite{mattis}:
\begin{equation}
\label{int}
f(k)=\int_{-\pi}^{\pi} \frac{dq}{2\pi}\,
\frac{U f(q)}{R_E(k)R_E(q)\left[{\mathcal E}(k,q)
-E\right]} \; ,
\end{equation}
where ${\mathcal E}(k,q)=\epsilon_\downarrow (k)+\epsilon_\downarrow (q) +
\epsilon_\uparrow(P-k-q)$, $P$ being the quasi-momentum of the trimer and
$\epsilon_\sigma(k)=2 t_\sigma (1-\cos k)$ the energy dispersions  of the two
components. Moreover $R_E(q)=[1+U I_E(q)]^{1/2}$, with~:
\begin{equation}
\label{I_E}
I_E(k)=\int \frac{dp}{2\pi}\, \frac{1}{{\mathcal E}(k,p) -E} \; .
\end{equation}
Eq.\ (\ref{int}) can be considered
as an eigenvalue problem ${\bf K}_E \cdot \bf f= \lambda \bf f$, where the
energy $E$  is fixed by the constraint $\lambda=1$.
\begin{figure}
\includegraphics[width = 0.7\columnwidth,keepaspectratio=true]{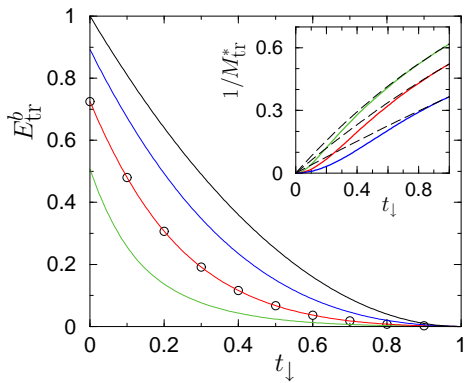}
\caption{Solid lines: Binding energy $E_\mathrm{tr}^b$ of a trimer (one light and two heavy fermions) 
as a function of the hopping ratio $t_\downarrow$ for
increasing values of the attraction strength $U=-2$ (bottom), $-4,-8, -\infty$
calculated from the exact solution of the 3-body problem. Open circles
represent DMRG calculations  in a chain of $L=100$
sites. \textsl{Inset}: Inverse effective mass of the trimer as a function of
the hopping ratio $t_\downarrow$  for increasing values of the attraction
strength $U=-2$ (top), $-4,-8$, shown by solid lines. At the symmetric
point $t_\downarrow=1$ the bound state disappears and the effective mass
reduces to the sum of the masses of constituents (dashed lines).}
\label{figEtrim}
\end{figure}
We solve this equation numerically for zero quasi-momentum $P=0$. The
binding energy $E_\mathrm{tr}^b$ of the trimer is related to the total energy
$E$ by $-E=E_\mathrm{pair}^b + E_\mathrm{tr}^b$, where $E_\mathrm{pair}^b=-2(1
+t_\downarrow)+\sqrt{U^2+4(1+t_\downarrow)^2}$ is the  binding energy of the
constituent pair \cite{2body}. In Fig.\ \ref{figEtrim} we plot the binding energy of the trimer as
a function of the mass asymmetry $t_\downarrow$ for increasing values of the
attraction $U$.
We see that $E_\mathrm{tr}^b$ vanishes at the symmetric point
$t_\downarrow=1$, in agreement with the Bethe Ansatz solution~\cite{takahashi}. As the mass
asymmetry increases the binding energy also increases until it saturates  at
$t_\down \rightarrow 0$. In this limit the function (\ref{I_E}) reduces to a
constant $I_E(k)=1/\sqrt{E(E-4)}$, implying that
Eq.\ (\ref{int}) has solution of the form $f(k)=\sin k$. Substituting this
into Eq.\ (\ref{int}) and integrating over momentum we  find $E=-U^2/(1-U)$. This gives
\begin{equation}\label{falikov}
 E_\mathrm{tr}^b(t_\down= 0)=\frac{U^2}{1-U}+2-\sqrt{U^2+4}\;,
\end{equation}
in agreement with our numerical solution. In particular, for infinitely strong
attraction, Eq.\ (\ref{falikov}) yields $E_\mathrm{tr}^b=1$,
showing that the trimer binding energy remains \textsl{finite} in contrast with the
pair binding energy  which is instead \textsl{divergent}.
Indeed, when the heavy particles are at neighboring sites, the light fermion can hop from one site to the other
without changing the interaction energy. Therefore the total energy gain is at most equal to $t_\up$.
In the strong coupling limit, $|U| \gg 1$, Eq.\ (\ref{int}) can be solved 
by the ansatz $f(k)=\sin k/R_E(k)$ yielding $E_\mathrm{tr}^b(U=-\infty)=(t_\downarrow-1)^2$, 
which is shown in Fig.\ \ref{figEtrim} with black line.

%


Let us now briefly discuss the effective mass $M_\mathrm{tr}^*$ of the trimer
which is defined by $1/M_\mathrm{tr}^*=\partial^2 E/\partial P^2$ evaluated at
$P=0$. The inverse effective mass is plotted in the inset of Fig.\
\ref{figEtrim}  as a function of the hopping rate $t_\downarrow$ and for
different values of the attraction strength. We see that the trimer becomes heavier as
$t_\down$ decreases or $|U|$ increases. At the breaking point, $t_\down=1$,
 the effective mass  reduces to
$(\sqrt{4(t_\downarrow+1)^2+U^2}+2)/4t_\downarrow$, corresponding to the sum
of the masses of a pair \cite{2body}, and of a heavy fermion. This quantity is plotted in the inset of Fig.\
\ref{figEtrim} with dashed lines.

\noindent\textit{Trimer gap.---} We now turn our attention to the effects of trimers at finite density. We first calculate the
trimer gap, namely the energy needed to break a single trimer. This is defined as
%
\begin{eqnarray}\label{trimergap}
& \Delta_\mathrm{tr} &=-\lim_{L  \rightarrow \infty} [ E_L(N_\up +1,N_\down+2)+E_L(N_\up ,N_\down) \nonumber\\
&&-E_L(N_\up +1,N_\down+1)-E_L(N_\up,N_\down+1)]\; ,
\end{eqnarray}
where $ E_L(N_\uparrow,N_\downarrow)$ is the ground state energy of a gas with
spin populations $N_\uparrow,N_\downarrow$ in a chain of size $L$. The limit in Eq.\ \eqref{trimergap} is taken assuming
$N_\sigma\to\infty$ with $n_\sigma\equiv N_\sigma/L$ being fixed. The trimer gap (\ref{trimergap}) is the generalization of the binding energy
at finite density, with the reference state being the many-body state $(N_\up,N_\down)$ rather than the vacuum $(0,0)$. 
We evaluate Eq.\ \eqref{trimergap} numerically via DMRG technique on
lattices of up to $L=160$ sites with open boundary conditions and perform
careful finite-size scaling in order to extract the thermodynamic limit
behavior. For equal masses, $t_\down=1$, our results are consistent with
$\Delta_\mathrm{tr}=0$ for any concentration, as expected from the Bethe
Ansatz solution.
For unequal masses, corresponding to $t_\down \neq 1$, the trimer gap (\ref{trimergap}) is finite only when the two concentrations 
 are commensurate, namely $n_\down=2 n_\up$. 

\begin{figure}[tb]
\includegraphics[width = 0.7\columnwidth,keepaspectratio=true]{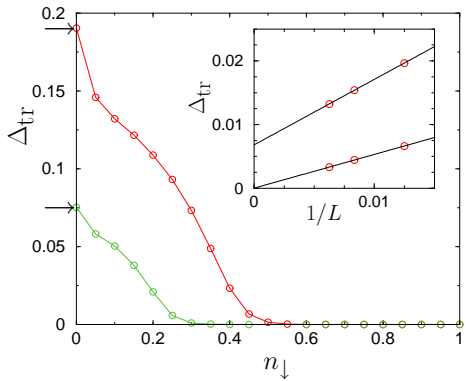}
\caption{Trimer energy gap plotted versus density $n_\down$ of the heavy component
 $(n_\up=n_\down/2)$ and for different values of the attraction $U=-2$ (bottom) and $U=-4$. 
The mass anisotropy is $t_\down=0.3$. The gap $\Delta_\mathrm{tr}$ reduces to the binding 
energy $E_\text{tr}^b$ at zero density (see arrows) and vanishes at a critical value of the density.
The data are obtained from finite size scaling after DMRG simulations with system sizes
$L=80,120,160$, assuming a linear dependence in $1/L$.  In the inset we show the scaling analysis for 
two different concentrations $n_\down=0.45$ and $n_\down=0.6$ with $U=-4$.} \label{fig_D3}
\end{figure}

Fig.\ \ref{fig_D3} shows
typical results for $\Delta_\mathrm{tr}$ at fixed $t_\down=0.3$ and
$n_\down/n_\up=2$. For vanishing density the trimer gap reduces to the binding energy $E_\mathrm{tr}^b$, 
shown in Fig.\ \ref{fig_D3} by arrows. As the density increases, $\Delta_\mathrm{tr}$
decreases and eventually \textsl{vanishes} at a critical concentration
$n_\down=n_\down^\mathrm{cr}$--- in a sharp contrast with the case of equal
densities, $n_\up=n_\down$, where the associated pairing gap is \textsl{always}
positive for any filling. In other words, the opening of the trimer gap is a non perturbative effect 
requiring finite coupling strength, or equivalently, low enough densities. This is consistent with our bosonization
approach \cite{bosonization}. 



\begin{figure}[htb]
\includegraphics[width = 0.80\columnwidth,keepaspectratio=true]{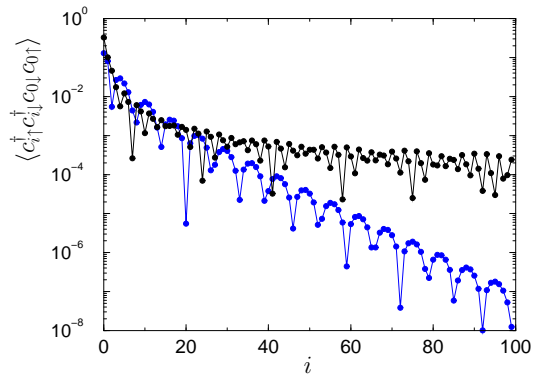}

\caption{Superconducting correlations as a function of the distance from the
center of the chain. The upper curve corresponds to  $n_\down=0.7$, where
the trimer gap is zero, cf. Fig.\ \ref{fig_D3}. The lower curve refers
to low density $n_\down=0.3$, where $\Delta_\mathrm{tr}>0$. 
The parameters used are $U=-4$, $t_\downarrow=0.3$ and
$L=200$ and the densities are commensurate, $n_\up=n_\down/2$. 
 Notice the
change from algebraic decay (upper curve) to exponential decay (lower curve).} \label{correlations}
\end{figure}

\noindent\textit{Correlation functions.---} 
The opening of the
trimer gap drastically affects the ground state properties of the gas, since
correlation functions of all operators breaking trimers fall off exponentially
rather than algebraically. In particular, the superconducting correlations
decay as~:
\begin{equation}
\langle c_{i\uparrow}^\dagger c_{i\downarrow}^\dagger
c_{j\downarrow}c_{j\uparrow}\rangle\propto
\frac{\exp(-|i-j|/\xi)}{|i-j|^\alpha} \cos(Q|i-j|)
\end{equation}
where the decay length $\xi \propto \Delta_\mathrm{tr}^{-1}$, $Q\equiv
|k_F^\up -k_F^\down|$ is the FFLO momentum, and $\alpha$ is a nonuniversal
number. 
Two-point correlations $\langle c^\dagger_{i\sigma} c_{j\sigma}
\rangle$ display similar behavior.
In Fig.\ \ref{correlations} we explicitly check this prediction by showing the
superconducting correlations in the gapped
$(n_\downarrow<n_\downarrow^\mathrm{cr})$ and gapless $(n_\downarrow >
n_\downarrow^\mathrm{cr})$ phases. We see that in the gapped phase the typical FFLO modulation is preserved,
as discussed in Ref.\cite{Batrouni2009,DasSarma2009}, but quasi long range order is lost.

\begin{figure}[htb]
\includegraphics[width = 0.7\columnwidth,keepaspectratio=true]{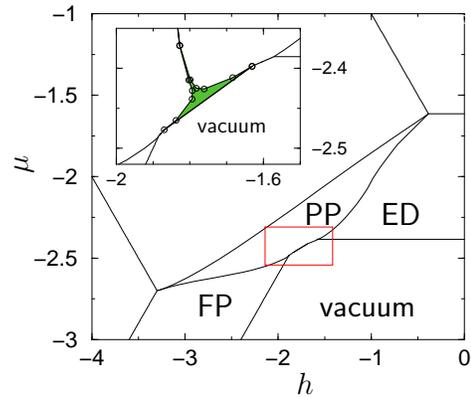}
\caption{Phase diagram for unequal masses obtained from DMRG simulations. Here
$t_\downarrow=0.3$ and $U=-4$. The novel line boundary between partially
polarized phase and vacuum is a consequence of the existence of $n$-body bound
states with $n>2$. Inset: a zoom-in of the low density region of the PP phase. The locus of commensurate 
densities $n_\up=n_\down/2$ is shown with green color. 
For clarity we only display the $h<0$ part of the phase diagram corresponding
to a majority of heavy ($\down$) fermions. The $h>0$ side is immediately
obtained by the particle-hole transformation $\mu\rightarrow -\mu+U,
h\rightarrow -h$. \label{diagram}  }
\end{figure}

\noindent\textit{Phase diagram.---} The presence of trimers and other bound
states induced by the mass asymmetry changes the topology of the grand
canonical phase diagram of the gas. The latter is obtained by replacing the
densities $n_{\up,\down}$ by two new variables, corresponding to the mean
chemical potential $\mu= \partial E / \partial (N_\up + N_\down)$ and  the
\textsl{effective} magnetic field $h=\partial E / \partial (N_\up - N_\down)$,
where $E$ is the ground state energy.
The evolution of the overall shape of the phase diagram with changing
$t_\down$ has been presented in Ref.\ \cite{DasSarma2009}. Here we concentrate
on the non-trivial changes due to the extra bound states induced by the mass
asymmetry, which were not discussed previously. To that end we work at fixed
values of $U$ and $t_\down$. We approximate the derivatives by
finite-difference formulas similar to Eq.\ \eqref{trimergap}, and  obtain the
phase diagram shown in Fig.\ \ref{diagram}. The partially polarized phase (PP)
corresponds to configurations with imbalanced spin populations $(1> n_\down >
n_\up >0)$. This phase is limited from the right by the phase of equal
densities (ED) $(n_\down = n_\up)$ and from below by the \textsl{fully
polarized} (FP) phase where the minority component is absent $(n_\up=0)$
\cite{note}.
For equal masses, $t_\down=1$ the three phases meet at a single point.
In presence of $n$-body bound states with $n>2$ this special point splits
into an extended line, corresponding to a direct boundary between PP phase and
vacuum. To see this, suppose there exists a bound state made of $p$
$\up$-fermions and $q$ $\down$-fermions, where $p$ and $q$ are non negative
integers. Since the system density is zero, the Taylor expansion
$E_{L=\infty}(p,q)= (p+q)\mu+(p-q)h$ becomes exact, yielding a straight line
in the $(h,\mu)$ plane associated to the bound state. The true phase boundary
$\mu=\mu_{vac}(h)$ with the vacuum is given by
\begin{equation}\label{vacuum}
\mu_{vac}=\mathrm{min}_{p,q} \frac{E_{L=\infty}(p,q)-(p-q)h}{p+q},
\end{equation}
resulting in a piecewise straight line, cf. Fig.\ \ref{diagram}. While for
equal masses the only states entering \eqref{vacuum} are a single
$\down$-fermion and a pair ($p=0$, $q=1$ and $p=q=1$, respectively), for
$t_\down \neq 1$ additional bound states appear, \textit{e.g.} trimers ($p=1$,
$q=2$), quadrimers ($p=1$, $q=3$), pentamers ($p=2$, $q=3$), etc.--- leading
to an extended PP-vacuum boundary, as shown in Fig.\ \ref{diagram}.
It is also instructive to consider the locus of $n_\up=n_\down/2$ on the phase
diagram. At low density $(n_\down<n_\down^\mathrm{cr})$, the trimer gap is
non-zero and the commensurate densities occupy a finite area of the $(h,\mu)$
plane. At higher density
$(n_\down> n_\down^\mathrm{cr})$, the energy gap closes and the locus shrinks
to a single line, as illustrated in the inset of Fig.\ \ref{diagram}. 
As the mass asymmetry increases, the trimer phase grows in size. Similar behavior 
is found for the gapped phases associated to all other bound states (not shown in Fig.\ \ref{diagram}).
We emphasize that in all the simulations reported above we observed a uniform ground state, apart from the usual Friedel oscillations
due to the open boundary conditions. Collapsed phases occur for  larger values of the ratio $|U|/t_\down$ \cite{Batrouni2009}. 

\noindent\textit{Relevance for experiments.---} Let us finally discuss how our  predictions can be tested in experiments with trapped Fermi gases. 
The asymmetric Hubbard model (\ref{aHubb}) can be directly implemented by using spin-dependent optical lattices \cite{mandel}. The presence of \textsl{shallow} harmonic  
 traps $V_\sigma^\textrm{ho}(z)=m\omega_{z \sigma}^2 z^2/2$ can be taken into account via local density approximation, starting from the homogeneous solution. Here $m$ is the atom mass and  $\omega_{z \sigma}$  the trapping frequencies. 
In order to form trimers, we require $T, \hbar \omega_ {z \sigma}\lesssim E_\mathrm{tr}^b$, $T$ being the temperature of the gas.
 For instance, consider a sample of $^6$Li atoms in a lattice with periodicity $d=250$ nm
and tunnelling rates $t_\uparrow=345 $nK and $t_\downarrow=0.2t_\uparrow$. Assuming 
$U=-4t_\uparrow$, from Fig.\ref{figEtrim} we obtain $E_\mathrm{tr}^b = 0.3 t_\uparrow=104$ nK, well within experimental reach.
The binding energy of trimers (Fig.\ \ref{figEtrim})  can be directly measured by rf spectroscopy \cite{esslinger, note2}. The 
effective mass (inset in Fig.\ref{figEtrim}) can be obtained by measuring the frequency of the dipole oscillations of the cloud \cite{PitaevskiiRMP2008}. For small displacements around the equilibrium position, the kinetic energy of trimers is quadratic, $P^2/2M^*_\mathrm{tr}$, and the latter is simply given by  $\omega_\textrm{dip}=\sqrt{(\omega_\uparrow^2+2\omega_\downarrow^2)m/M^*_\mathrm{tr}}$. 
To enter the trimer phase at \textsl{finite density}, the above condition becomes more stringent, namely $T, \hbar \omega_ {z \sigma}\lesssim \Delta_\mathrm{tr}$. Moreover both components must be degenerate, implying $T \lesssim E_{F \uparrow},E_{F \downarrow} $, where $E_{F \sigma}$ are the Fermi energies in the absence of interaction. This sets a lower bound on the values  of the densities at the center of the trap. For the above choice of parameters,  the best trade-off occurs around  $n_\downarrow(0)=2 n_\uparrow(0) \sim 0.3/d$ yielding  $T, \hbar \omega_ {z \sigma}\lesssim 60$ nK. 
Notice that to have an extended trimer phase, the two spin populations should be tuned close to the commensurate point $N_\uparrow=N_\downarrow/2$, otherwise phase separation in shells will occur \cite{Orso2007,Drummond2007}.
Finally, the suppression of the superconducting correlations (Fig.\ref{correlations}), signalling the emergence of the trimer phase, can in principle be detected using \textsl{interferometric} techniques, as discussed in Ref.\cite{castin}.

\textit{Concluding}, we have shown that 1D attractive fermions with unequal masses  form trimers and other more exotic  bound states.
Differently from pairs, these states can only open a gap at low density or, equivalently, strong interactions. In the gapped phase 
FFLO superconducting correlations are exponentially suppressed.  
The properties of trimers in vacuum and at finite densities are experimentally accessible with ultra-cold atoms. Numerical simulations were performed using the
ALPS libraries \cite{ALPS}.


%
%
%
%
%
%


\begin{thebibliography}{99}
\bibitem{note2} 
Notice that three-body losses for trimers are strongly suppressed due to Pauli principle.
\end{thebibliography}
\end{document}